\documentclass[%
 reprint,
 amsmath,amssymb,
 aps,
]{revtex4-2}

\usepackage{graphicx}
\usepackage{dcolumn}
\usepackage{bm}

\begin{document}


\title{Hawking radiation as a manifestation of spontaneous symmetry breaking}



\author{Ivan Arraut}
\email{ivan.arraut@usj.edu.mo}
\affiliation{
University of Saint Joseph,\\
Estrada Marginal da Ilha Verde, 14-17, Macao, China
}%


\date{\today}

\begin{abstract}
We demonstrate that the black hole evaporation can be modelled as a process where one symmetry of the system is spontaneously broken continuously. We then identify three free-parameters of the system. The sign of one of the free-parameters, governs whether the particles emitted by the black-hole are fermions or bosons. The present model explains why the Black Hole evaporation process is so universal. Interestingly, this universality emerges naturally inside certain modifications of gravity.
\end{abstract}

\maketitle


\section{Introduction}

Black holes are highly compact objects, generating very strong gravitational fields. They concentrate all their mass inside the gravitational radius $2GM$, which becomes to be an event horizon for the case of spherical symmetry \cite{1, 111}. For Kerr-black holes (rotating), as well as charged black-holes, some modifications for this expression are done \cite{1, 111, 2, 211}. In any case, once an object crosses the event horizon of a black-hole, it can never escape from the huge gravitational attraction experienced at such scales. This is a classical perspective of a black hole. Quantum mechanically however, it is known that the black holes emit particles in the form of a spectrum of radiation. This was the seminal discover of Hawking \cite{3}. Although Hawking's original formulation focused on the use of Bogoliubov transformations, several other methods were developed, including path integrals, etc \cite{4}. The Hawking's mechanism (radiation), brought with itself a huge theoretical problem, namely, the black-hole information paradox \cite{5}. The paradox suggests that since the black holes emit the particles in the form of radiation, then they come out without any information from the past. This is called lost of unitarity. Unitarity is one of the fundamental conditions satisfied in Quantum Mechanics \cite{6, 611}. By the date there is no a universally accepted solution for the black-hole information paradox, although it is strongly suspected that the solution must be connected with the concepts related to the holographic principle \cite{7, 711}. Considering this important information problem, it becomes highly priority to analyze the Hawking radiation from different perspectives. This possibility is open now with the outcome of analogue models, some of them including holographic principles \cite{8, 811}. In previous papers, the black-hole evaporation by using analogue models has been done, including an analysis from the perspective of neural networks \cite{9, 10}. In this paper we demonstrate that the Hawking radiation is equivalent to the spontaneous breaking of some symmetry of the system. The general idea is that the Black Hole starts with a state with some specific mass, angular momentum and charge ($M_1$, $L_1$ , $Q_1$). Although there are many possible internal Black Hole configurations able to reproduce these combinations (degenerate vacuum states), at the moment when the radiation is emitted, only one of those configurations is allowed. This is equivalent to a system selecting one specific vacuum state. Subsequently, after the emission of radiation, the system will be in a different state with $M_2$, $L_2$ and $Q_2$. Since there are many different states with internal configurations able to satisfy this condition, the system again selects one specific vacuum configuration. Each time that the system selects a particular vacuum state, then the symmetry is spontaneously broken and then radiation is emitted. In this paper we construct a model where a scalar field represents the particle number for the particles emitted by the Black-Hole. The scalar field Lagrangian then contains a potential term with a scalar field expansion of the order of quadratic, cubic and quartic order. The system then has three free-parameters. The relation between the different parameters, changes the vacuum configuration and then the behavior of the particles. In particular, the sign of the parameter related to the cubic interaction term for the particle-number field, defines whether the particles evaporating are fermions or bosons. In the limit, when this parameter vanishes, there is no distinction between fermions and bosons. We interpret this result in the language of Black-Holes. The paper is organized as follows: In Sec. (\ref{BHE}), we we revise the standard formulation of the black-hole evaporation process. In Sec. (\ref{Basics}), we revise the basic concepts related to the mechanism of spontaneous symmetry breaking. In Sec. (\ref{Basics2}), we formulate the Black-Hole evaporation process, or equivalently, the particle creation process, as a consequence of breaking spontaneously the symmetry under exchange of configurations (exchange of particles) due to the presence of the black-hole. In Sec. (\ref{Basics3}), we explain how the curvature effects appear from the particle Lagrangian proposed in order to explain the Hawking radiation process. Finally, in Sec. (\ref{Conclude}), we conclude.

\section{Standard formulation of the Black Hole evaporation}   \label{BHE}

\begin{figure}[h!]
\centering
\includegraphics[scale=0.14]{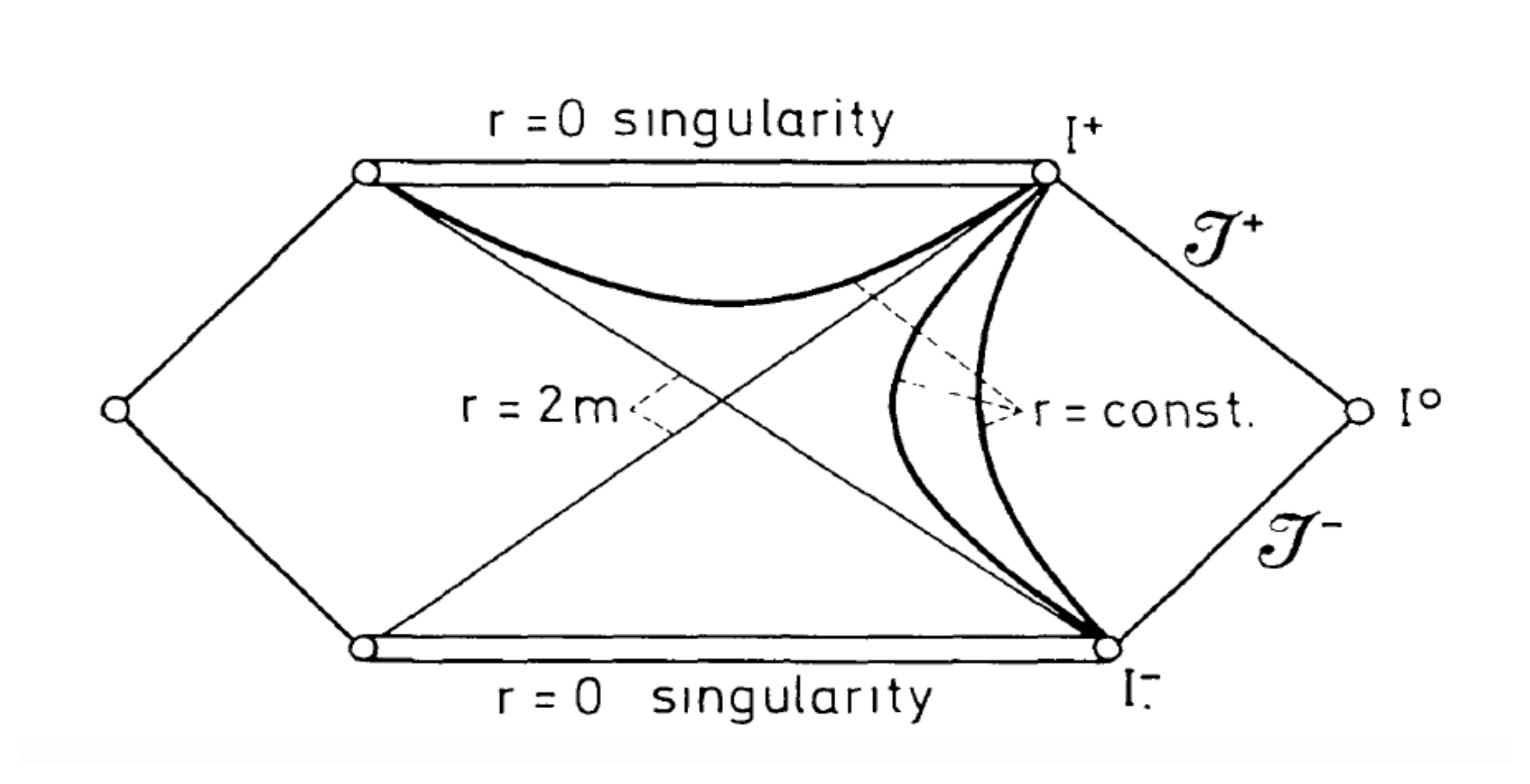} 
\caption{The Penrose diagram for the Schwarzschild geometry in GR as it is showed in \cite{3}. The past-null infinity $({J}^-)$ of the diagram, represents all the possible events happening before the formation of the black-hole. On the other hand, the future null infinity, represents all the possible events happening after the formation of the black-hole.\label{fig1}}
\end{figure}

The black hole evaporation process is a quantum effect, derived originally inside a semiclassical approach \cite{3}. It is semiclassical because the derivation is not carried inside a full formulation of quantum gravity. Instead, the calculation works inside the classical background of General Relativity with a quantum field moving around. The emission of particles coming from a Black Hole can be explained when we observe a Penrose diagram as it is shown in the figure (\ref{fig1}). We can then imagine an observer located at the infinity with respect to the event horizon and observing the dynamic of a quantum scalar field, which represents the matter content perceived at that point in spacetime. The quantum field at that point can be expanded as 

\begin{equation}   \label{field1}
\phi(x, t)=\sum_{\bf p}\left(f_{\bf p}\hat{b}_{\bf p}+\bar{f}_{\bf p}\hat{b}_{\bf p}^+\right).    
\end{equation}
This expansion contains all the information of the quantum field. The original argument of Hawking, suggests that the emission of particles emerge when an observer compares two vacuum states. The first vacuum state is defined before the black-hole is formed and then it can be defined as

\begin{equation}   \label{fvs}
\hat{b}_{\bf p}\vert\bar{0}>=0.    
\end{equation}
Then the expansion (\ref{field1}) corresponds to the field expansion before the formation of the black-hole (past-null infinity). However, after the formation of the black-hole, the vacuum configuration changes and the quantum scalar field in eq. (\ref{field1})) now must be expanded in terms of a different set of modes as follows

\begin{equation}   \label{field2}
\phi(x, t)=\sum_{\bf p}\left(p_{\bf p}\hat{a}_{\bf p}+\bar{p}_{\bf p}\hat{a}_{\bf p}^++q_{\bf p}\hat{c}_{\bf p}+\bar{q}_{\bf p}\hat{c}_{\bf p}^+\right).
\end{equation}
Both expansions, (\ref{field1}) and (\ref{field2}), must have the same information. From the result (\ref{field2}), we define a new vacuum state as $\hat{a}_{\bf p}\vert0>=0$. This vacuum emerges after the formation of the black-hole. In eq. (\ref{field2}), the modes $q_{\bf p}$, together with the operators $\hat{c}_{\bf p}$, are defined at the event horizon. These modes hide behind the event horizon of the black-hole. Then the standard calculation suggests that we have to compare the vacuum state defined in eq. (\ref{fvs}) with the vacuum state defined with respect to the operator $\hat{a}_{\bf p}$ in eq. (\ref{field2}). These can be done via Bogoliubov transformations. Following the arguments of Hawking in \cite{3}, we can find that the relation between the modes under discussion, respect the following relation

\begin{equation}   \label{realbogo}
\hat{a}_{\bf p}=u_{{\bf p}, {\bf p'}}\hat{b}_{\bf p'}-v_{{\bf p}, {\bf p'}}\hat{b}_{\bf p'}^+. 
\end{equation}
Then we can see why the vacuum perceived after the formation of the black-hole is not empty even if it is devoid of particles before the formation of the black-hole. The density of particles emitted by the black hole, is defined by the following expression

\begin{equation}   \label{mama2}
<\bar{0}\vert\hat{n}_{\bf p}^a\vert\bar{0}>=\vert v_{{\bf p}, {\bf p'}}\vert^2.
\end{equation}
The arguments of Hawking, based on the dynamics of a quantum field interacting with the black-hole, showed that the amount of particles emitted by a black-hole, follow the following statistical distribution

\begin{equation}   \label{Statistics}
<\bar{0}\vert\hat{n}_{\bf p}^a\vert\bar{0}>=\frac{\Gamma_{{\bf p}, {\bf p'}}}{e^{\frac{2\pi\omega}{\kappa}}\pm1}.    
\end{equation}
The negative sign is taken if the particles escaping the Black-Hole are bosons, while the positive sign is taken if the emitted particles are fermions. In eq. (\ref{Statistics}), $\Gamma_{{\bf p}, {\bf p'}}$ represents the portion of particles going inside the Black-Hole \cite{3}. The Bogoliubov coefficients then satisfy the following relation

\begin{equation}
\vert u_{{\bf p}, {\bf p'}}\vert=e^{\frac{\pi\omega}{\kappa}}\vert v_{{\bf p}, {\bf p'}} \vert. 
\end{equation}

\section{Spontaneous symmetry breaking: Basic concepts}   \label{Basics}

\begin{figure}[h!]
\centering
\includegraphics[scale=0.4]{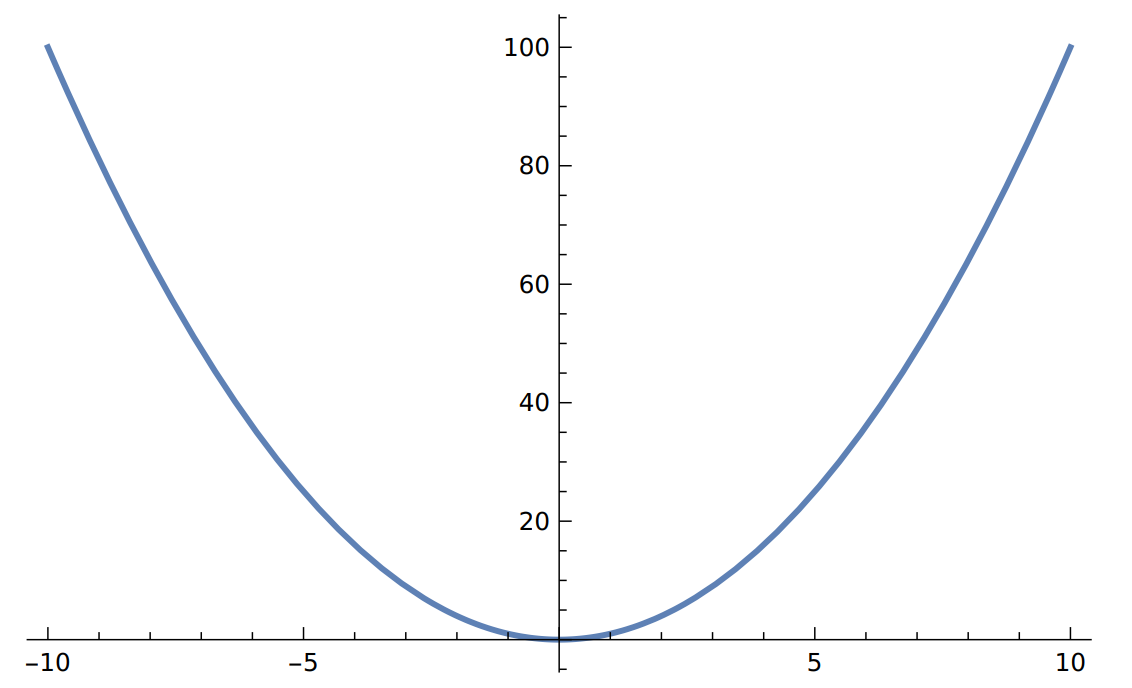} 
\caption{Potential with one stable equilibrium point. This is the shape of the potential emerging when the parameter $m$ satisfies the condition $m^2>0$ in eq. (\ref{Supert}). In this case $m$ represents the mass of the field $\phi$ and the symmetry of the system, is not spontaneously broken. \label{fig2}}
\end{figure}

In this section, we will introduce the basic notions of spontaneous symmetry breaking \cite{Nambu1, Nambu2, Nambu5, Nambu3, Nambu4, 11}, with the intention of applying these concepts at the moment of analyzing the particle creation process of a black-hole. Spontaneous symmetry breaking occurs when the ground state of a system violates a symmetry which is still satisfied by the Langrangian of the same system. This occurs for certain combinations of the free-parameters of the Lagrangian. A typical case is the standard linear $\sigma$-model (here simplified), where a Lagrangian of the form \cite{11}

\begin{equation}
\pounds=\frac{1}{2}(\partial_\mu\phi)(\partial^\mu\phi^{*})-V(\phi, \phi^*), 
\end{equation}
with the potential defined as

\begin{equation}   \label{Supert}
V(\phi, \phi^*)=\frac{1}{2}m^2\phi^*\phi+\frac{\lambda}{4!}(\phi\phi^*)^2,
\end{equation}
with the two free-parameters $m^2$ and $\lambda$, can be defined. The ground state of this system is obtained from the condition $\partial V/\partial\phi=0$. If $m^2>0$, then we get the trivial result $\vert\phi_0\vert=0$. In these circumstances, this trivial ground state satisfies all the symmetries of the Lagrangian. We can then say that the symmetry is not yet broken spontaneously. This also means that the ground state represents a stable equilibrium as it appears in the figure (\ref{fig2}).

\begin{figure}[h!]
\centering
\includegraphics[scale=0.7]{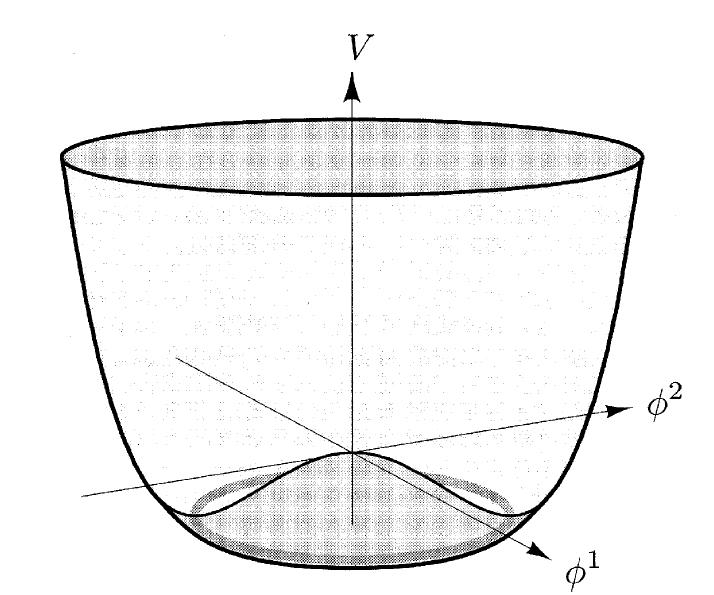} 
\caption{The potential (\ref{Supert}) for the case where the symmetry of the system is spontaneously broken. This occurs when $m^2<0$. For this case, there exists a multiplicity of ground states and the system selects one of them arbitrarily, depending on the direction of a external perturbation. \label{fig3}}
\end{figure}

However, when $m^2<0$, we cannot interpret $m$ as the mass anymore, and for these cases, the equilibrium condition $\partial V/\partial\phi=0$, gives two possible results. The first result gives $\vert\phi_0\vert=0$, but this is now an unstable equilibrium condition. This means that the system does not feel comfortable anymore staying at this point and instead, it will find another equilibrium condition representing stability. The new equilibrium condition, corresponds to the second solution of $\partial V/\partial\phi=0$, which gives

\begin{equation}
\vert \phi_0\vert^2=-\frac{m^2}{2\lambda}=a^2,    
\end{equation}
At the quantum level, we treat the field $\phi$ as a quantum field and then we conclude that the vacuum expectation value of the quantum field, is given by 

\begin{equation}
<o\vert\phi^2\vert0>=a^2.    
\end{equation}
It has been demonstrated that with a decomposition $\phi=\phi_1+i\phi_2$, the potential defined in eq. (\ref{Supert}) corresponds to the one appearing in the figure (\ref{fig3}) \cite{11, Ryder}.
From this example, we can see that when $m^2>0$, the stable vacuum condition is $<0\vert\phi\vert0>=\phi_0=0$ and it is unique. On the other hand, when $m^2<0$, the stable vacuum condition is now $<0\vert\phi\vert0>=\phi_0=a$ and it is not unique. Yet still the system has to select one among all the possible equivalent ground states. In the next section, we will see that the relevant dynamical quantum field for the case of black-holes, is the particle number operator $\hat{n}_{\bf p}$, with a zero vacuum expectation value before the formation of the black-hole (no spontaneous symmetry breaking at this point). After the formation of the black-hole, one symmetry is spontaneously broken and then the vacuum expectation value of $\hat{n}_{\bf p}$ is different from zero. Following the standard formulation, when $m^2<0$, once the system selects a ground state, this ground state violates certain symmetries of the Lagrangian. If we define all the symmetry transformations of the potential (\ref{Supert}) as $U(g)$, then the violation of some of these group of symmetries means that the transformations \cite{Ryder}

\begin{equation}
U(g)\phi_0\neq\phi_0,    
\end{equation}
do not keep invariant the ground state in general. The symmetry transformation could be more explicitly defined as $U=e^{i\hat{G}\alpha}$, with $U$ being a unitary operator and $\hat{G}$ being a Hermitian operator. Yet still, there remains some subgroup of symmetries (generators) which still represent a symmetry for the ground state. The relevant expression here is

\begin{equation}
U(h)\phi_0=\phi_0.    
\end{equation}
Here we can again define some symmetry generators $\hat{H}$ which are unbroken. In summary, when the symmetry of a system is spontaneously broken, then the ground state (stable equilibrium condition) is degenerate (multiple ground states) and once the system selects one of the possible ground states, the selected state violates some of the symmetries still satisfied by the original Lagrangian. Additionally, the ground state takes a non-trivial value (different from zero).

\section{Black Hole evaporation as a consequence of spontaneous symmetry breaking}   \label{Basics2}

The black hole evaporation effect can be also obtained if we take the particle number operator $\hat{n}^a_{\bf p}$ as quantum scalar field. Then we define its Lagrangian as the one containing a kinetic term and and potential term with the field expansion done up to fourth-order, without omitting the cubic term in the field expansion. Working in this way, we have a Lagrangian of the form

\begin{equation}   \label{KGextended}
\pounds=\frac{1}{2}\partial^\mu\hat{n}^a_{\bf p}(\omega)\partial_\mu\hat{n}^a_{\bf p}(\omega)-V(\hat{n}^a(\omega)),
\end{equation}
with the partial derivative taken with respect to the frequency $\omega$, namely, $\partial_\mu=\partial_\omega=\frac{\partial}{\partial\omega}$. The potential term in eq. (\ref{KGextended}) is 

\begin{equation}   \label{KGextended2}
V(\hat{n}_{\bf p})=\frac{1}{2}m^2\hat{n}_{\bf p}^2+\frac{\beta}{3}\hat{n}^3_{\bf p}+\frac{\lambda}{4}\hat{n}^4_{\bf p}.
\end{equation}
In eq. (\ref{KGextended2}) we have omitted the index $a$ for simplicity. Although in principle, the ground state for the potential is defined by solving the equation $\partial V/\partial\hat{n}=0$; we cannot assume this to be the correct result in this case because the spacetime curvature forces the field $\hat{n}_{\bf p}$ to have a kinetic term at its most stable configuration. Still, it can be proved that the symmetry of the system under exchange of internal configurations consistent with the Black-Hole entropy (exchange of particles), is spontaneously broken when $m^2<0$. In addition, the signature of the parameter $\beta$ tells us whether the particles evaporating are bosons or fermions. In other words, we have two solutions for this case. The first solution corresponds to the boson statistics and the second one corresponds to the fermion statistics. Both statistics are identical to the ones defined in eq. (\ref{Statistics}) and equally obtained inside the standard formalism due to Hawking. The black hole naturally emits both types of particles. When the surface gravity of the black-hole tends to zero, the statistics of fermions and bosons emitted by the black-hole, have a similar behavior. However, when the surface gravity of the Black Hole increases, then more bosons than fermions are emitted \cite{3}. From these explanations, it is clear that the parameter $\beta$, must be related to the surface gravity as we will demonstrate in a moment. Now we can now proceed to analyze the behavior of the solution of the Euler-Lagrange equations obtained from the Lagrangian defined in eq. (\ref{KGextended}).  

\subsection{Emission of particles}

Here we will explore the solutions for the field equations, corresponding to the Lagrangian in eq. (\ref{KGextended}). The system in general, has the tendency for looking for some equilibrium configuration, but it is unable to reach such condition and instead it emits particles at each instant of selection of some configuration. The solution to the field equations, obtained from the Lagrangian (\ref{KGextended}), is given by 

\begin{equation}   \label{sigmoig-like}
\hat{n}^a=\frac{A}{e^{-\gamma \omega}\pm1},    
\end{equation}
with the corresponding parameters $m^2=-\gamma^2$, $\beta=3\gamma^2/A$ and $\lambda=-2\gamma^2/A$, as they are defined inside the potential (\ref{KGextended2}). The sign of $\beta$ defines whether the emitted particles are bosons or fermions. If we compare eq. (\ref{sigmoig-like}) with eq. (\ref{Statistics}), then $\gamma$ takes negative values in this case. Note that the parameters $m$, $\beta$ and $\lambda$ do not depend on the sign taken by $\gamma$. After doing the corresponding comparisons, then $\gamma$ is related to the surface gravity as
\begin{equation}   \label{surfgamma}
\gamma=-\frac{2\pi}{\kappa}.    
\end{equation}
Then the surface gravity is equivalent to $\kappa=-2\pi/\gamma$. It is a trivial task to notice that $A=\Gamma_{\bf p, p'}$ inside the present comparison. Note that since $m^2$ is negative in this case, the symmetry under exchange of particles (keeping the same total mass, angular momentum and charge) is spontaneously broken. Then a black-hole is just an unstable system, permanently trying to find its ground state configuration but it never reaches it until it evaporates completely. 

\subsection{The connection of $\beta$ with the particle statistic}

For seeing the connection between the sign of $\beta$ and the statistic followed by the particles emitted, we analyze the Euler Lagrange equations obtained from eq. (\ref{KGextended}), but considering dimensionless coupling constants. In such a case, we get

\begin{equation}   \label{Femnexap}
(\partial^\mu\partial_\mu+\bar{m}^2)n^a_R+\bar{\beta}(n^a_R)^2+\bar{\lambda}(n^a_R)^3=0.
\end{equation}
Here the sub-index $R$ makes reference to the fact that we are dealing with a dimensionless equation which will give a dimensionless solution. We can convert easily the dimensionless solution toward the full solution (\ref{sigmoig-like}). For getting the fermionic statistic, we need to satisfy the set $\bar{\lambda}=-2$, $\bar{\beta}=3$ and $\bar{m}^2=-1$. On the other hand, for getting the bosonic statistic, the same set of parameters are valid, except the value of $\bar{\beta}$, which for the bosonic case takes the value $\bar{\beta}=-3$. It is a trivial task to demonstrate the following expressions

\begin{equation}   \label{expressions}
m^2=\bar{m}^2\gamma^2, \;\;\;\;\;\beta=\frac{\bar{\beta}\gamma^2}{A},\;\;\;\;\;\lambda=\frac{\bar{\lambda}\gamma^2}{A}.    
\end{equation}
Then the change in sign of the dimensionless parameter $\bar{\beta}$, affects the sign of $\beta$, and then the statistic following by the emitted particles depends on the signature taken by $\beta$. Combining eq. (\ref{expressions}), with (\ref{surfgamma}), we get

\begin{equation}
\beta=\frac{4\pi^2\bar{\beta}}{A\kappa^2}.
\end{equation}
From this expression, it is clear that when $\bar{\beta}\to0$, then $\kappa\to0$. Then more properly, when $\bar{\beta}=0$, equal amount of fermions and bosons are emitted by the black-hole. However, if instead we keep $\bar{\beta}=\pm3$ in agreement with the paragraphs following eq. (\ref{Femnexap}), then the condition for getting equal amount of fermions and bosons is $\beta\to\pm\infty$, with the positive sign corresponding to the fermionic statistic. 

\subsection{Symmetry analysis of the phenomena}

Taking into account the theory developed in the section (\ref{Basics}), we can see that the ground (vacuum) state for the black-hole is characterized by the value of the particle number operators $\hat{n}_{\bf p}$. The symmetry of the black-hole corresponds to the invariance under internal particle distributions, as far as the mass, angular momentum and charge of the black-hole does not change. For simplicity, let's consider the scenario of a Schwarzschild black-hole, where there is no angular momentum and there is no charge. For this case, for a given mass $M$, the ground state of the black-hole, is invariant under exchange of particles or internal changes of configuration. Let's define this symmetry with the unitary operator $U(g)$, with a generator $\hat{G}$, this time indicating exchange of particles internally. Before the formation of the black-hole, the ground state is trivial and then the vacuum expectation value  satisfies $<0\vert\hat{n}^a_{\bf p}\vert0>=0$. This result is obtained from the potential (\ref{KGextended2}) when $m^2>0$, $\beta>0$ and $\lambda>0$. However, when $m^2<0$, $\beta>0$ and $\lambda<0$, the vacuum configuration changes and then the trivial ground state $<0\vert\hat{n}^a_{\bf p}\vert0>$ is now unstable and then the system evolves towards a new stable vacuum state configuration with $<0\vert\hat{n}^a_{\bf p}\vert0>\neq0$, which is degenerate in agreement with the black-hole entropy, which counts the number of possible internal configurations. The new vacuum expectation value of the particle number operator, is then given by eq. (\ref{sigmoig-like}). Under these conditions, the black-hole selects a specific vacuum configuration and then the ground state does not respect anymore the symmetry under exchange of configurations. It is under these circumstances that the grounds state, represented by the vacuum expectation value $<0\vert\hat{n}^a_{\bf p}\vert0>\neq0$, becomes non-trivial and then the particle emission process continues. This process is continuous and it only finishes when the black-hole evaporates completely (ignoring any possible quantum gravity effect at the Planck scale). Let's define the particle exchange (Hermitian) operator as $\hat{P}$. Then the symmetry operation is defined by the unitary operator $U(p)=e^{-i\hat{P}\alpha}$. Before the formation of the black-hole, for the trivial vacuum state, the particle exchange operation will not change the vacuum condition. In this case then

\begin{equation}
U(p)<0\vert\hat{n}_{\bf p}^a\vert0>=<0\vert\hat{n}_{\bf p}^a\vert0>=0.
\end{equation}
On the other hand, after the formation of the black-hole, the symmetries under exchange of particles are lost at the vacuum level due to spontaneous symmetry breaking and then we have in general 

\begin{equation}
U(p)<0\vert\hat{n}_{\bf p}^a\vert0>\neq<0\vert\hat{n}_{\bf p}^a\vert0>.
\end{equation}
The effect of $U(p)$ over the ground state is to map it towards another equivalent but different ground state. The symmetry under exchange of particles, which is connected with the black-hole entropy, when broken spontaneously, is then the mechanism behind the emission of particles of a black-hole. In the following section, we explain the generic character of the Lagrangian (\ref{KGextended}).

\section{Curvature effects appearing from the particle Lagrangian}   \label{Basics3}

The first impression coming from the Lagrangian (\ref{KGextended}), is that gravity is apparently absent and then all the calculations would represent a simple coincidence between the results obtained by Hawking in eq. (\ref{Statistics}) and the one obtained in this paper in eq. (\ref{sigmoig-like}). However, these types of coincidences do not exist and here we will prove that in fact, gravity appears implicit inside the Lagrangian defined in eq. (\ref{KGextended}). The Lagrangian of a standard scalar field moving along a flat spacetime (without gravity), is defined as 

\begin{equation}
\pounds=\frac{1}{2}\partial^\mu\phi(x)\partial_\mu\phi(x)-\bar{m}^2\phi^2(x).    
\end{equation}
Here $\bar{m}$ is the mass of the Quantum field moving along the flat spacetime. The vacuum state of this Quantum field is simply $\phi(x)=0$, if we ignore the residual vacuum energy coming from the ground state of the Quantum harmonic oscillator \cite{11}. Now let's introduce gravity over this system such that the Quantum field moves now along a curved spacetime with minimal coupling. In such a case, the Lagrangian takes the form

\begin{equation}   \label{Curvlagaga}
\pounds=\frac{1}{2}\sqrt{-g}\left(g^{\mu \nu}\phi_{,\mu}(x)\phi_{,\nu}(x)-m^2\phi^2(x)\right)
\end{equation}
Here the gravity effects emerge from the deviations of the metric with respect to the Minkowski spacetime. Although the spacetime curvature generated by a Black-Hole is very large, for an initial explanation, we can apply perturbation theory over the spacetime metric, in order to analyze how the terms appearing on the potential (\ref{KGextended2}) emerge. Perturbative theory takes the small deviations of the metric $g_{\mu \nu}$ with respect to Minkowski as $g_{\mu \nu}\approx \eta_{\mu \nu}+h_{\mu \nu}$. Here $\eta_{\mu \nu}$ is the Minkowski spacetime, while $h_{\mu \nu}$ is the perturbation around Minkowski. In this way, $\sqrt{-g}\approx 1+\frac{1}{2}h+\frac{1}{8}h^2-\frac{1}{4}h_{\mu\nu}^2+...$ up to second order, with $h=0$ in vacuum and $h\propto T$ when there is a source term $T\propto \phi^2(x)$ at the ground state. We can also make similar statements for the case $h_{\mu \nu}\propto T_{\mu\nu}$ since at this point we are only concerned about proportionality relations. Then the Lagrangian near the ground state (ignoring kinetic terms) now becomes

\begin{equation}   \label{pert}
\pounds\approx -\frac{1}{2}\left(1+\frac{1}{2}h+\frac{1}{8}h^2-\frac{1}{2}h_{\mu\nu}^2\right)\bar{m}^2\phi^2. 
\end{equation}
If we expand the Lagrangian (\ref{Curvlagaga}) by considering the previous comments and the result (\ref{pert}), then it is evident that terms of different orders on the field $\phi(x)$ will emerge if we take into account that $h\propto T\propto\phi^2(x)$. This also means that terms of different orders in the particle number operator $\hat{n}_{\bf p}$ will emerge, considering that naively $\hat{n}_{\bf p}\propto \phi^2(x)$, given the fact that the scalar fields are linear functions of the annihilation and creation operators. With these arguments, the Lagrangian (\ref{pert}), generate terms of the form

\begin{equation}   \label{Lag2}
\pounds=a\hat{n}_{\bf p}+b\hat{n}_{\bf p}^2+c\hat{n}_{\bf p}^3+...    
\end{equation}
The expansion include higher order terms at the non-linear level, which increase in relevance. If we compare eq. (\ref{Lag2}) with eq. (\ref{pert}), then obviously certain terms in the expansion in eq. (\ref{Lag2}), would correspond to the terms in the potential (\ref{KGextended2}) in a direct way. There will be other terms in the expansion difficult to compare, unless a re-summation between terms emerge at the event horizon level. Yet still, we can see that each term in eq. (\ref{KGextended2}) can be reproduced from the Einstein-Hilbert expansion no matter what. These type of re-summation methods appear in massive gravity in order to eliminate an undesirable ghost at the non-linear level \cite{NLMG}, \cite{NLMG2}, \cite{NLMG3}. Since massive gravity converges to General Relativity when the gravitational field is strong, then the amount of particles emitted from the event horizon of a black-hole in General Relativity is the same amount of particles emitted from the event horizon of a black-hole inside the non-linear theory of massive gravity, as it was demonstrated in \cite{MyHawking}, \cite{MyHawking2}. Based on this interesting aspect for the black-holes, it is important to realize that although the metric expansions developed in \cite{NLMG}, \cite{NLMG2}, \cite{NLMG3} were done thinking on a massive theory of gravity (non-linear), still the same formalism is general in the sense that we can use it for analyzing certain aspects of gravity. In \cite{NLMG} is illustrated how the deviations with respect to Minkowski can be represented in a non-linear theory of gravity as $g_{\mu\nu}=\eta_{\mu\nu}+h_{\mu \nu}=H_{\mu\nu}+Special\;\;terms$. Here the special terms refer to those terms carrying out gravitational degrees of freedom by using the St\"uckelberg trick \cite{Stu}. In this way, at the end of the calculations, it was demonstrated that if we want to find the source term of the Einstein equations, it can be calculated from a potential term containing quadratic, cubic and fourth order terms in the metric (the same order corresponds to the particle number operator) \cite{NLMG3}

\begin{equation}   \label{potentialmassive}
U_{source}=U_2+\alpha_3U_3+\alpha_4U_4.
\end{equation}
This potential contains three free-parameters which can be paired with three free-parameters of the Einstein-Hilbert action after considering the field equations \cite{NLMG3}

\begin{equation}   \label{FE}
G_{\mu\nu}=-m^2X_{\mu\nu},    
\end{equation}
with $X_{\mu\nu}=\frac{\delta U}{\delta g^{\mu\nu}}-\frac{1}{2}Ug_{\mu\nu}$. What is important to remark here is that there is a direct connection between the series expansion of the standard Einstein-Hilbert action and the potential expansion defined in eq. (\ref{potentialmassive}) through the Euler-Lagrange equations, which give us the field equations in (\ref{FE}). In eq. (\ref{potentialmassive}), since each term $U_n\propto g_{\mu\nu}^n\propto \phi^{2n}\propto n_{\bf p}^n$, then we have a direct correspondence between the potential defined in eq. (\ref{potentialmassive}) and the potential proposed in eq. (\ref{KGextended2}). Then the Hawking radiation effect is so general, that the form of the Lagrangian reproducing it from eq. (\ref{KGextended2}) appears in theories intending to generate source terms with degrees of freedom being able to move through an event horizon. The result is generic and it explains why the Hawking radiation obeys the bosonic/fermionic statistics of a black-body. In other words, if the Lagrangian (\ref{KGextended}) had a different potential term instead of (\ref{KGextended2}), then the statistics followed by the spectrum of Black-hole would change dramatically. This can be seen if we evaluate the Euler-Lagrange equations over eq. (\ref{KGextended}). 
   
\section{Conclusions}   \label{Conclude}

In this paper we have proved that it is possible to model the black-hole evaporation process as a mechanism of continuous symmetry breaking, where the black hole is permanently selecting some specific vacuum configuration, forcing then the system to emit particles. This means that a black-hole, having a degenerate vacuum state in agreement with its entropy, will select a ground state among all the possibilities consistent with its mass, angular momentum and charge. Once this occurs, the system emits radiation, decaying towards a new configuration with different values of charge, angular momentum and mass. The process then continues until the black-hole evaporates completely. This means that the black-hole is permanently looking for one stable configuration but it never reaches it and that's why the Hawking radiation process emerges. Finally, we have also identified and analyzed some free parameters for the field theory which is able to model the Black-Hole evaporation process. Inside the free-parameters, the signature of $\beta$ or the signature of its dimensionless counterpart $\bar{\beta}$, determine whether the particles evaporating are fermions or bosons. The Lagrangian proposed for modelling the scalar field moving around a Black-Hole, is naturally coupled to gravity because the higher order terms or self-interaction terms are precisely consequence of this coupling. Finally, we have demonstrated the generic character of the Hawking radiation by showing that similar potential forms emerge from theories reproducing the propagation of gravitational degrees of freedom through black-hole event horizons \cite{NLMG3}. In fact, by using the standard and generic method where a non-linear formulation of gravity can include a source term (depending on the scalar field), a potential of the form (\ref{KGextended2}) emerges. This should not be a surprise because in massive gravity the degrees of freedom propagating in addition to the tensorial (spin-2) component is a scalar (spin-0) component. Then in gravity in general, the propagation of a scalar field in the presence of curvature effects is a generic process. In other formulations of gravity, additional terms on the potential defined in eq. (\ref{KGextended2}) might appear, depending on the corrections to the evaporation process proposed by the different theories. Then for example, at the Quantum regime \cite{F1, F2, F3}, naturally certain modifications to the Hawking approach are required and this will bring as a consequence modifications of the Lagrangian (\ref{KGextended}). These modifications might appear as quantum corrections to the Lagrangian (\ref{KGextended}) and they can generate additional spontaneous symmetry breaking patterns \cite{QFTF}. These ideas will be discussed further in future papers. The proposed formulation is the ideal one for for analyzing the black-hole information paradox by using as a starting point the Lagrangian proposed in eq. (\ref{KGextended}). In the past, other authors did some attempts, formulating the black-hole evaporation process via the spontaneous symmetry breaking of certain symmetries \cite{Moffat} or by using neural network model approaches inspired in Bose-Einstein condensates \cite{10}. However, these approaches are not as generic as the one proposed in this article. In future papers we will be exploring the black-hole information paradox from this perspective.

\nocite{*}


\end{document}